\documentclass[compactaffiliation]{Interspeech}

\interspeechcameraready

\title{The Risks and Detection of Overestimated Privacy Protection\\in Voice Anonymisation}

\author[affiliation={1}, equalcontribution]{Michele}{Panariello}
\author[affiliation={2}, equalcontribution]{Sarina}{Meyer}
\author[affiliation={3}]{Pierre}{Champion}
\author[affiliation={4}]{Xiaoxiao}{Miao}
\author[affiliation={1}]{Massimiliano}{Todisco}
\author[affiliation={2}]{Ngoc Thang}{Vu}
\author[affiliation={1}]{Nicholas}{Evans}

\affiliation{}{EURECOM}{France}
\affiliation{Institute for Natural Language Processing}{University of Stuttgart}{Germany}
\affiliation{}{Inria}{France}
\affiliation{}{Singapore Institute of Technology}{Singapore}

\email{michele.panariello@eurecom.fr, sarina.meyer@ims.uni-stuttgart.de}

\usepackage{comment}

\usepackage{xcolor}
\newcommand{\red}[1]{#1}

\newcommand{\asvsys}[1]{\text{ASV}_\text{#1}}
\newcommand{\traindata}[1]{T_{\text{#1}}}
\newcommand{\evaldata}[1]{E_{\text{#1}}}

\newcommand{\eerval}{EER\textsubscript{val}}
\newcommand{\eertest}{EER\textsubscript{test}}
\DeclareMathOperator{\ohnn}{OHNN}
\newcommand{\secref}[1]{Section~\ref{#1}}
\newcommand{\tabref}[1]{Table~\ref{#1}}
\newcommand{\figref}[1]{Figure~\ref{#1}}

\usepackage{tabularray} 

\usepackage{floatrow}
\newfloatcommand{capbtabbox}{table}[\captop][\Xhsize] 
\begin{document}

\maketitle


\begin{abstract}
Voice anonymisation aims to conceal the voice identity of speakers in speech recordings. 
\red{Privacy protection} is usually estimated from the difficulty of using a speaker verification system to re-identify the speaker post-anonymisation. Performance assessments are therefore dependent on the verification model as well as the anonymisation system. There is hence potential for \red{privacy protection} to be overestimated when the verification system is poorly trained, perhaps with mismatched data. 
In this paper, we demonstrate the insidious risk of overestimating anonymisation performance and show examples of exaggerated performance reported in the literature.
For the worst case we identified, performance is overestimated by 74\% relative.
We then introduce a means to detect when performance assessment might be untrustworthy and show that it can identify all overestimation scenarios presented in the paper.
Our solution is openly available as a fork of the 2024 VoicePrivacy Challenge evaluation toolkit.
\end{abstract}

\SetTblrInner{rowsep=0.9pt} 

\section{Introduction}
\label{sec:introduction}

Voice anonymisation is the task of processing a speech recording to conceal the voice identity of the speaker while retaining utility --- with utility being defined according to some downstream task~\cite{introducing_vpc}.
State-of-the-art solutions operate upon multiple features extracted from the input speech signal. Typically, they encode at least the desired information pertaining to the downstream task (e.g.\ spoken and emotional content).
Voice conversion is then applied to synthesise an output speech signal which is devoid of the original voice identity but which contains the same desired information as the input.  The anonymised output then contains the voice belonging to a so-called \textit{target speaker}.
Anonymisation is successful if an attacker cannot establish the voice identity of the original speaker.

\red{Privacy protection} is typically measured by simulating attacks with an automatic speaker verification (ASV) system which has the aim of linking anonymised data to the original speaker. Performance is inversely proportional to ASV reliability.
Simulated attacks should be as strong as possible so that anonymisation performance is measured in a worst-case sense. This is usually achieved by training the ASV model using utterances anonymised by the same anonymisation system under evaluation. Within the scope of the VoicePrivacy Challenge (VPC)~\cite{vpc22_taslp, vpc24_eval_plan}, this is referred to as the \emph{semi-informed} attack model. The VoicePrivacy Attacker Challenge~\cite{attacker_challenge} was launched recently to foster the development of stronger attacks to improve the reliability of performance estimates. 
At the time of writing, the semi-informed attack model and evaluation recipe described in~\cite{vpc24_eval_plan} remains the de-facto standard.



Estimates of \red{privacy protection} reflect both that of the anonymisation system as well as the ASV system used for evaluation. Performance can be overestimated if the ASV system performs sub-optimally for reasons other than the anonymisation itself. For example, previous work~\cite{pierre_legendary_thesis} showed that some system design choices might prevent the ASV system from learning discriminative cues from anonymised data, resulting in exaggerated estimates of anonymisation performance.  
They are due not to strong anonymisation, 
but rather to the use of a suboptimal ASV attack model. 
The inadvertent overestimation of performance can be challenging to avoid; 
contrary to the design of reliable ASV systems, strong anonymisation calls for \emph{poor} ASV performance. Poor performance can be achieved relatively easily, perhaps due to inadequate training choices or even design oversight.
As a result, and other than to protect scientific integrity, there is comparatively less incentive to optimise an ASV model used for the evaluation of anonymisation systems. This raises challenges in  benchmarking; results depend on the degree to which each ASV attack model is optimised with respect to the anonymisation system under evaluation.

In this paper, we report an investigation of potential evaluation pitfalls 
and propose what is, to the best of our knowledge, the first solution for their detection. We show that exaggerated \red{privacy protection} estimates can result from a mismatch  between the distributions of anonymised test data and that used to train the ASV attack model.  
We demonstrate the pitfalls using artificial scenarios and then show similar trends in the evaluation of practical anonymisation systems reported in the literature.
Once the distribution mismatch is addressed, we show that levels of anonymisation performance can, in some cases, be much lower than those reported. 
Finally, 
we show that the proposed approach to detection is effective in protecting against untrustworthy performance estimates.

\begin{figure}[t]
  \centering
    \caption{Illustration of the mismatched evaluation procedure (best viewed in colour). $\text{s}_1$ and $\text{s}_2$ are voice anonymisation systems from $P$. Matched: $\text{s}_1 = \text{s}_2$. Full mismatch (\secref{subsec:complete_mismatch}): $\text{s}_1 \neq \text{s}_2$, different colours. Partial mismatch (\secref{subsec:partial_mismatch}): $\text{s}_1 \neq \text{s}_2$, same colour, different shapes. Hidden mismatch (\secref{subsec:hidden_mismatch}): $\text{s}_1 \neq \text{s}_2$, same colour, different borders.}
  \includegraphics[width=1\columnwidth]{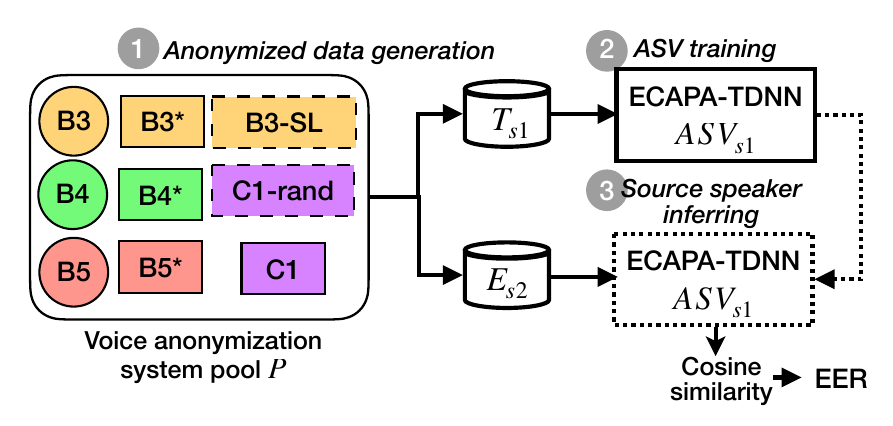}
  \label{fig:overal}
\vspace{-1.5em}
\end{figure}

\section{Anonymisation methods}\label{sec:methods}
We report an evaluation of anonymisation performance for a set of scenarios involving some form of mismatch between the anonymisation system under test and that used to generate ASV training data. In this section we describe the set of systems used in our experiments.  They include 
3 VPC 2024 baseline systems~\cite{vpc24_eval_plan}. For reasons which will become apparent later, we also used one more system based on self-supervised learning (SSL) and orthogonal householder neural networks (OHNN)~\cite{ohnn}.

Baseline \textbf{B3}~\cite{stttttts} uses a transformer-based system~\cite{peng2022branchformer} to produce a phonetic transcript of the input audio. An acoustic model is used with a prosodic representation of the input and a GAN-generated~\cite{liu2019wasserstein} speaker embedding to produce an anonymised Mel spectrogram which is converted to a waveform using a HiFi-GAN vocoder~\cite{kong_hifi-gan_2020}.
Baseline \textbf{B4}~\cite{nac} extracts a representation of the spoken content from the input utterance in the form of semantic tokens.  Using the
EnCodec~\cite{encodec} encoder, a speaker representation in the form of acoustic tokens is extracted from a second, distinct utterance which contains the voice of the target speaker. The two sets of tokens are modeled in auto-regressive fashion to generate a new set of anonymised acoustic tokens which are decoded to a waveform using the \mbox{EnCodec} decoder.
Baseline \textbf{B5}~\cite{pierre_legendary_thesis} is based upon the
quantisation of wav2vec 2.0~\cite{wav2vec2} bottleneck features, which are concatenated with the F0 curve and a one-hot
representation of the target speaker identity. Voice conversion is performed using a HiFi-GAN vocoder.
The SSL-OHNN system, henceforth referred to as \textbf{C1}, uses a HuBERT-based soft content encoder~\cite{hubertsoft} to extract the spoken content and an ECAPA-TDNN model~\cite{the_one_and_only_ecapa} to extract a speaker embedding from the original utterance. Anonymisation of the speaker embedding is
performed using an OHNN while voice conversion is performed directly using a HiFi-GAN vocoder.

We evaluate anonymisation performance according to the usual VPC 2024 policy~\cite{voicepat, vpc24_eval_plan} and as shown in \figref{fig:overal}. The evaluation of system S is performed using two sets of anonymised LibriSpeech data~\cite{librispeech}: (a)~the LibriSpeech test data~($\evaldata{S}$)~\cite{vpc24_eval_plan}; 
(b)~the LibriSpeech \textit{train-clean-360} set of data~\cite{librispeech} which is used by the attacker to train an ASV system using similarly anonymised data.
The datasets are anonymised in one of two different ways: (a)~at the \textit{utterance level}, in which each input utterance is anonymised towards a different target speaker; (b)~at the \textit{speaker level} in which, for all utterances produced by the same original speaker, anonymisation is performed towards the same target speaker.

For each anonymisation system S, the attacker trains a new ASV model $\asvsys{S}$ using anonymised training data $\traindata{S}$. A data partition corresponding to 10\% of $\traindata{S}$ is set aside and used for validation during training.
For all experiments reported in this paper, $\asvsys{S}$ is an ECAPA-TDNN model~\cite{the_one_and_only_ecapa}.
\red{Its use for privacy evaluation is mandated by the VPC evaluation plan~\cite{vpc24_eval_plan} and makes our results comparable to others reported in the anonymisation literature.}
Speaker embeddings are extracted from the final hidden layer for anonymised enrollment and test utterances.  The speaker of the former is either the same  (a target trial) or different  (a non-target trial) to that of the original test utterance (pre-anonymisation). 
Embeddings are scored using a cosine distance backend.
As per~\cite{vpc24_eval_plan}, \red{privacy protection} is expressed in terms of the equal error rate (EER) for $\asvsys{S}$ when evaluated using $\evaldata{S}$.  The higher the EER, the better the anonymisation: the attacker has more difficulty to reidentify the original speaker. 
An EER of 50\% would indicate full anonymisation: the attacker cannot do better than to guess at random whether the speaker of the enrollment and test utterances is the same.
This attack is referred to as \textit{semi-informed}: the attacker knows the system S used to generate $\evaldata{S}$
and uses S to produce $\traindata{S}$ with which to train $\asvsys{S}$, as well as to anonymise the enrollment utterances.
Readers are referred to~\cite{vpc24_eval_plan, evaluating_voice_conversion_based_privacy_protection} for details.

\section{Experiments}
We first 
show that differences between the anonymisation systems used to generate anonymised evaluation data $\evaldata{S}$ and that used by the adversary for ASV training $\traindata{S}$ can lead to \red{privacy protection} overestimation. We then show similar findings when the two anonymisation systems differ only in terms of a single module.
Next, we show that estimates of performance can be lower still if the attacker 
instead trains
the ASV system in a scenario subtly different to the usual \textit{semi-informed} attack. In all of these cases, there is a \emph{mismatch} between the two anonymisation systems. Last, we propose a technique to detect such mismatches with a view to mitigating the overestimation of anonymisation performance.


\begin{table}[t]
    \centering
    \caption{Results for full mismatch scenarios (EER, \%). The diagonal corresponds to the matched condition, the remaining values to the full mismatch.}
    \vspace{-0.5em}
    \begin{tblr}{c|ccc}
    \hline[1.3pt]
        \SetCell[r=2]{c} Attacker   & \SetCell[c=3]{c}{Evaluation Data}    &                   & \\
                                    & $\evaldata{B3}$                      & $\evaldata{B4}$   & $\evaldata{B5}$ \\
    \hline[1pt]
        $\asvsys{B3}$   & \textbf{27.05} & 44.23          & 44.40        \\
        $\asvsys{B4}$   & 36.99          & \textbf{29.82} & 41.67         \\
        $\asvsys{B5}$   & 37.28          & 38.72          & \textbf{34.34} \\
    \hline[1.3pt]
    \end{tblr}
    \label{tab:exp_complete_mismatch}
\vspace{-1em}
\end{table}

\subsection{Full mismatch} \label{subsec:complete_mismatch}
Under \textit{full mismatch} conditions, the two anonymisation systems are completely different. As illustrated in \figref{fig:overal}, system $\text{s}_1$ is used by the attacker to generate ASV training data $\traindata{s\textsubscript{1}}$. System $\text{s}_2$ is used to produce anonymised evaluation data $\evaldata{s\textsubscript{2}}$. 
We performed experiments using all possible pairings among systems B3, B4 and B5. 
Results are presented in \tabref{tab:exp_complete_mismatch}.
Each row reflects performance for  $\asvsys{s\textsubscript{1}}$ trained using $\traindata{s\textsubscript{1}}$, while columns correspond to evaluation data $\evaldata{s\textsubscript{2}}$.
Shown in bold face along the diagonal are results for matched conditions where $\text{s}_1 = \text{s}_2$. 
This is the usual \emph{semi-informed} attack scenario. 
Off-diagonal results correspond to the conditions of interest where $\text{s}_1 \neq \text{s}_2$.
\red{While this setup does not represent a realistic evaluation scenario, the implications of mismatched anonymisation systems remain unclear and have not been previously explored.}

Our results show that the impact of such a mismatch can be substantial.
An
EER of 27\% is produced when anonymised evaluation data and ASV training data are both generated using the same B3 system. However, the EER increases to 44\% EER (near random guess) when the ASV system is trained using data generated using B4 or B5. Results are similar for B4 for which the EER increases from 30\% (matched) to 42\% (B5). For B5 the effect is less pronounced, yet not insignificant.
While the above is an extreme, unrealistic example, we show next the potential for similar performance overestimates even when the differences between each anonymisation system is more subtle, including real examples from the literature.


\begin{table}[t]
    \centering
    \caption{Results for the partial mismatch scenario.
    \vspace{-0.5em}
    }
    \begin{tblr}{c|c|c}
    \hline[1.3pt]
    Evaluation data & Attacker & EER (\%) \\
    \hline[1.3pt]
    \SetCell[r=2]{c} $\evaldata{B3}$ & $\asvsys{B3}$  & 27.05 \\
                                  & $\asvsys{B3*}$ & 28.11 \\
    \hline
    \SetCell[r=2]{c} $\evaldata{B4}$ & $\asvsys{B4}$  & 29.82 \\
                                  & $\asvsys{B4*}$ & 35.83 \\
    \hline
    \SetCell[r=2]{c} $\evaldata{B5}$ & $\asvsys{B5}$  & 34.34 \\
                                  & $\asvsys{B5*}$ & 43.71 \\
    \hline[1.3pt]
    \end{tblr}
    \label{tab:exp_partial_mismatch}
    \vspace{-1em}
\end{table}

\subsection{Partial mismatch} \label{subsec:partial_mismatch}
We evaluate the performance of an anonymisation system under a more realistic, \emph{partial mismatch} scenario in which the anonymisation systems differ only in terms of a single module.
Experiments were performed using the following system variants:
\textbf{B3*} in which 
the HiFi-GAN is replaced with a \mbox{BigVGAN} vocoder~\cite{lee2023bigvgan}; \textbf{B4*} in which
the EnCodec decoder is replaced with the Vocos~\cite{siuzdak2024vocos} vocoder trained with character-level conditioning~\cite{panariello24_spsc};
\textbf{B5*} which uses a factorised time delay neural network (\mbox{TDNN-F})~\cite{tdnn-f} bottleneck feature extractor instead of the wav2vec 2.0~\cite{wav2vec2} encoder.\footnote{System B5* is identical to B6 in~\cite{vpc24_eval_plan}. Denoted here as B5* for consistency as a \emph{variant} of system B5.} 
The experimental setup is unchanged, but this time involves the evaluation of $\evaldata{B3}$ using either $\asvsys{B3}$ or $\asvsys{B3*}$, $\evaldata{B4}$ using either $\asvsys{B4}$ or $\asvsys{B4*}$, and $\evaldata{B5}$ using either $\asvsys{B5}$ or $\asvsys{B5*}$.

Results are presented in \tabref{tab:exp_partial_mismatch}. 
For evaluation data $\evaldata{B3}$, the difference in performance is negligible, likely because the use of only a different vocoder has no consequential impact upon anonymisation behaviour. 
For $\evaldata{B4}$
the impact is more substantial, with a difference of 6\% in EER for $\asvsys{B4}$ and $\asvsys{B4*}$.
The character conditioning mechanism of B4* likely has a greater impact on anonymisation.
For $\evaldata{B5}$ the difference in EER is even more pronounced, with a difference of almost 10\%.
The use of a different feature extractor early in the pipeline 
results in almost completely different systems (cf.\ results in \tabref{tab:exp_complete_mismatch}).

Even a partial mismatch between anonymisation systems can still induce 
substantially exaggerated performance estimates.
While even this scenario might easily be avoided, there are practical examples in the literature.
In~\cite{ohnn} (C1) and the system proposed in~\cite{liping_and_I_had_a_3_hour_conversation_about_this_one}, the attacker is assumed to retrain one of the system modules, thereby inducing a partial mismatch. 
In~\cite{xinyuan24_spsc}, $\evaldata{S}$ is formed by mixing data generated using two different anonymisation systems, and the attacker is assumed not to know with which system each test utterance is anonymised; since the ASV model cannot be matched to both systems, there is again a partial mismatch.
In all of these works, evaluation is reportedly performed under the VPC-defined \textit{semi-informed} attack model. This observation shows that 
its definition is inadequate and that differences in interpretation
can lead to questionable performance comparisons.
This finding calls for the design of an automatic method to estimate the reliability of results derived using an ASV system $\asvsys{S}$ regardless of the attack definition. We propose one such method in \secref{subsec:detect_mismatch}.

\begin{figure*}[t]
    \centering
    \includegraphics[width=0.9\linewidth]{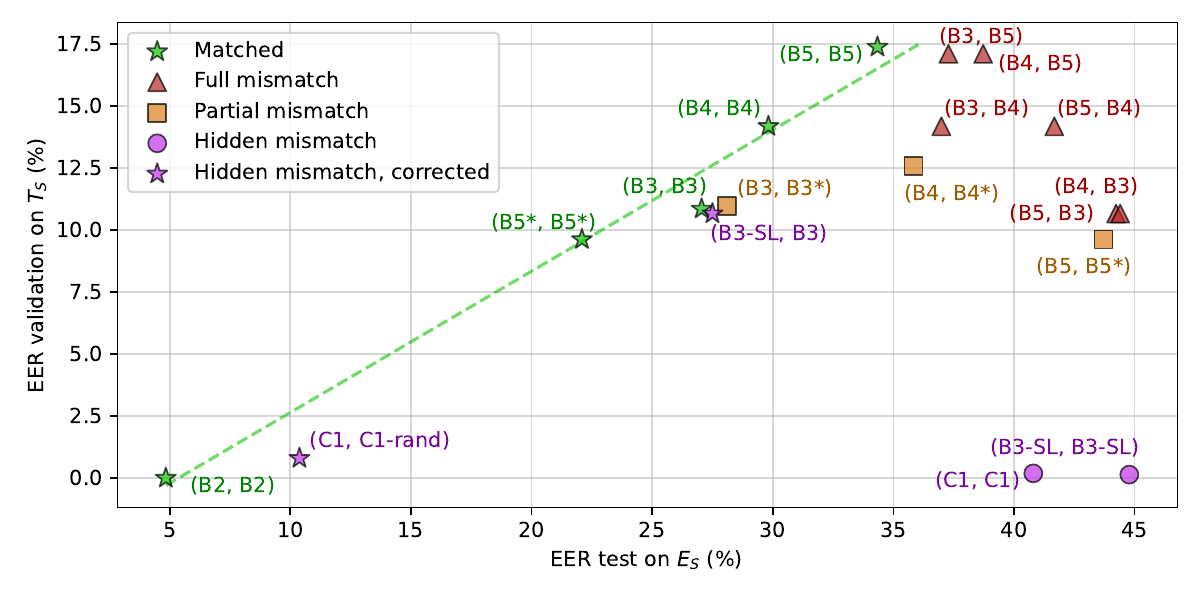}
      \vspace{-0.8em}
    \caption{Values of \eerval{} plotted against \eertest{} for all considered systems, denoted as ($\evaldata{S}$, $\traindata{S}$) pairs. The green dashed line is the regression line of the \textit{``Matched''} systems. Systems falling below this line have a potentially mismatched evaluation.
          \vspace{-1em}
      \vspace{-0.2em}
      }
    \label{fig:eerval_vs_eertest}
\end{figure*}

\subsection{Hidden mismatch} \label{subsec:hidden_mismatch}


As we now show, some mismatches can be more insidious.
We start with the case
of system \textbf{C1}~\cite{ohnn} for which results are shown to the left in \tabref{tab:exp_hidden_mismatch}. 
When evaluation is performed using $\asvsys{C1}$ and $\evaldata{C1}$ we obtain an EER of 41\%. 
However, we found that the EER can be reduced very substantially to 10\% (a 74\% relative decrease) through a trivial retraining of the ASV system.
The retrained system, denoted $\asvsys{C1-rand}$, 
stems from the anonymisation of each utterance within $\traindata{C1-rand}$ using an embedding extracted for a speaker selected at random from the \textit{\mbox{LibriTTS} train-other-500} dataset~\cite{libritts} rather than that extracted for speakers selected using the OHNN-based target speaker selection strategy.
We now illustrate why such a training approach results in a more effective attack.

The OHNN-based target speaker selection strategy is a completely deterministic function $\ohnn (\mathbf{x})$ applied to a speaker embedding $\mathbf{x}$.
It is well-known~\cite{drift, drift2-electric_boogaloo} 
that, if the target speaker selection strategy is not sufficiently random, then utterance-level anonymisation behaves similarly to speaker-level anonymisation.
This is because, if two embeddings $\mathbf{x}_1$ and $\mathbf{x}_2$ are extracted from utterances produced by the same speaker,
then $\mathbf{x}_1 \approx \mathbf{x}_2$. It follows then that $\ohnn (\mathbf{x}_1) \approx \ohnn (\mathbf{x}_2)$, hence speaker-level, rather than utterance-level anonymisation.

It has been shown previously that speaker-level anonymisation of $\traindata{S}$ results in a weaker attack~\cite{shamsabadi_differentially_2023}, yet a thorough explanation for why is lacking.
We now offer such an explanation.
\red{If speaker $A$ in $\traindata{S}$ is consistently anonymised towards 
target speaker $B$, $\asvsys{S}$ may simply learn to associate speaker $B$’s voice characteristics with label $A$. As a result, it may rely on the $B\Rightarrow{}A$ mapping instead of learning speaker-discriminative features that persist despite anonymisation. Consequently, $\asvsys{S}$ is likely to fail on $\evaldata{S}$, which includes different speakers and different anonymisation mappings.}
This is the case for C1 since 
the attacker retrains the OHNN, thereby resulting in a different transformation $\ohnn (\mathbf{x})$, the learning of unreliable cues, and hence overestimated anonymisation performance.
By using a random target speaker selection strategy which results in $\traindata{C1-rand}$ being anonymised at the utterance level, $\asvsys{C1-rand}$ will no longer learn unreliable cues. This results in the learning of other, more reliable cues, hence the lower EER.

To provide empirical evidence in support of these arguments, we conducted a set of experiments using a variant of B3, denoted B3-SL, which performs speaker-level rather than utterance-level anonymisation.  
We used B3-SL to create $\evaldata{B3-SL}$ and $\traindata{B3-SL}$, both anonymised at the speaker level but with different target speaker mappings.
As shown to the right in \tabref{tab:exp_hidden_mismatch}, the resulting $\asvsys{B3-SL}$ system produces an EER of almost 45\% when assessed using 
$\evaldata{B3-SL}$. We then repeated the evaluation using $\evaldata{B3-SL}$ and $\asvsys{B3}$ trained using 
data anonymised at the utterance level and obtained a greatly-reduced EER of 27\%.
$\asvsys{B3-SL}$ provides an overestimation of \red{privacy protection} as a result of being trained using data anonymised at the speaker level.
These findings show that the evaluation of anonymisation systems performed using the same system as that used in generating $\evaldata{S}$ may not always provide reliable performance estimates. Such hidden mismatches can be challenging to mitigate.

\begin{table}[t]
    \centering
    \caption{Results for the hidden mismatch scenario (EER, \%).
    \vspace{-0.5em}
    }
    \begin{tblr}{cc|cc}
    \hline[1.3pt]
    \SetCell[c=2]{c}{$\evaldata{C1}$} & & \SetCell[c=2]{c}{$\evaldata{B3-SL}$} \\
    $\asvsys{C1}$ & $\asvsys{C1-rand}$ & $\asvsys{B3-SL}$ & $\asvsys{B3}$ \\
    \hline
    40.80 & 10.38 & 44.78 & 27.31 \\
    \hline[1.3pt]
    \end{tblr}
    \label{tab:exp_hidden_mismatch}
    \vspace{-0.5em}
\end{table}

\subsection{Detecting mismatches} \label{subsec:detect_mismatch}



The overestimation of \red{privacy protection} can be attributed to $\asvsys{S}$ \textit{overfitting} to the data distribution of $\traindata{S}$ due either to the use of mismatched anonymisation systems (Sections~\ref{subsec:complete_mismatch} and~\ref{subsec:partial_mismatch}) or the use of ineffective training strategies (\secref{subsec:hidden_mismatch}). This causes $\asvsys{S}$ to underperform on $\evaldata{S}$.
In the following we propose one solution to detect overfitting 
using validation data~\cite{tibshirani_my_old_friend}. Our hypothesis is that differences in EER for validation and test data derived using an attacking ASV system can serve as an indication of overfitting. 
The validation data is the same 10\% of $T_s$ used for ASV training (see Section~2). In contrast to $\evaldata{S}$, the validation data contains the same speakers as $T_s$.
We design a new ASV protocol for the validation data.
For each speaker, 5 utterances are used for enrollment. Those remaining are reserved for ASV trials. On average, there are 8 trials per speaker, 3 of which are target trials. $\asvsys{S}$ is trained in the usual way and then applied to the validation data in addition to the usual evaluation data $\evaldata{S}$ resulting in a pair of performance estimates, \eerval{} and \eertest{}.
We performed this experiment using systems B3, B4 and B5 under full, partial and hidden mismatch conditions. To provide more data points, we also used the B5 variant B5* in addition to the less competitive B2 baseline~\cite{mcadams}.

Results are illustrated in \figref{fig:eerval_vs_eertest}, where the x and y axes correspond to \eertest{} and \eerval{} respectively. Green stars correspond to matched scenarios for 2024 VPC baseline systems. Red triangles correspond to full mismatch (\secref{subsec:complete_mismatch}). Orange squares correspond to partial mismatch (\secref{subsec:partial_mismatch}). Purple circles indicate hidden mismatch, while purple stars correspond to the same scenarios with corrected mismatch (\secref{subsec:hidden_mismatch}). Annotations signify the systems involved in each evaluation, e.g.\ (B4,B5) corresponds to $\evaldata{B4}$ being used for evaluation of $\asvsys{B5}$.
For all cases, \eerval{} is substantially lower than \eertest{}: this is on the account of the overlap between speakers in the validation set and the training set $\traindata{S}$.
The gap between \eerval{} and \eertest{} is more pronounced for mismatched scenarios. For instance, in the case of matched (B3,B3), the \eertest{} of $27\%$ falls to an \eerval{} of $11\%$, a relative drop of $60\%$. However,
for the full mismatched case of (B4,B3)
an \eertest{} of $44$\% falls to an \eerval{} of $10$\%, a relative drop of $76\%$.
The trend is consistent; for mismatched scenarios the gap between \eertest{} and \eerval{} is higher than that for matched scenarios.

Also plotted in \figref{fig:eerval_vs_eertest} is a regression line computed over the \eertest{}, \eerval{} point pairs  corresponding to matched scenarios, all of which we assume to be well-evaluated reference cases.
Nearly all other points, all corresponding to mismatched scenarios, fall below this line suggesting that the difference between \eerval{} and \eertest{} serves as an indicator of how \mbox{well-trained} or well-matched is $\asvsys{S}$ to the evaluation of anonymisation system S: the greater the gap, 
the higher the chance of there being a mismatch between training and evaluation data, and therefore a less reliable estimation of anonymisation performance. 
We also note that cases of hidden mismatch are the farthest from the regression line. However, corrected counterparts are closer to the regression line, indicating 
more reliable performance estimation.

We argue that the community should consider the adoption of such detection techniques for the evaluation of anonymisation systems so as to reduce the risk of overestimating performance. To encourage work in this direction, we have integrated our approach into a publicly available fork of the 2024 VPC evaluation toolkit\footnote{\url{https://github.com/DigitalPhonetics/voice-privacy-evaluation}} which includes the automatic computation of \eerval{} within the validation step.


\section{Conclusions}
In this paper we demonstrate the risk of overestimating anonymisation performance. 
Using several state-of-the-art anonymisation approaches, we show that the risk stems from mismatches between the data used to train the speaker verification system employed for evaluation and the anonymised data under test.  We demonstrate the risk with artificially introduced mismatch and that similar, hidden mismatches also afflict the evaluation of practical systems reported in the literature leading to exaggerated reports of performance.
Based upon the comparison of results derived for test and validation sets, our method to detect overestimated performance is effective in identifying all examples reported in the paper.
Consequently, we advocate
for the adoption of such detection techniques within the community and for future editions of the VoicePrivacy Challenge to protect the trust in performance estimates. 
Future work should extend our analysis to other sources of mismatch and their detection.

\ifinterspeechfinal
\section{Acknowledgements}
This work is funded by the Deutsche Forschungsgemeinschaft (DFG, German Research  Foundation) – Project: Multilingual Controllable Voice Privacy (VoiPy) - Project number 533241795. 
\else
\fi

\bibliographystyle{IEEEtran}
\bibliography{mybib}

\end{document}